\documentclass[sigconf]{acmart}

\AtBeginDocument{%
  }

\copyrightyear{2025}
\acmYear{2025}
\setcopyright{rightsretained}
\acmConference[RecSys '25]{Proceedings of the Nineteenth ACM Conference on Recommender Systems}{September 22--26, 2025}{Prague, Czech Republic}
\acmBooktitle{Proceedings of the Nineteenth ACM Conference on Recommender Systems (RecSys '25), September 22--26, 2025, Prague, Czech Republic}\acmDOI{10.1145/3705328.3748105}
\acmISBN{979-8-4007-1364-4/2025/09}

\usepackage{subcaption}

\newcommand{\squishlist}{
 \begin{list}{$\bullet$}
  { \setlength{\itemsep}{0pt}
     \setlength{\parsep}{3pt}
     \setlength{\topsep}{3pt}
     \setlength{\partopsep}{0pt}
     \setlength{\leftmargin}{1.5em}
     \setlength{\labelwidth}{1em}
     \setlength{\labelsep}{0.5em} } }

\newcommand{\squishlisttwo}{
 \begin{list}{$\bullet$}
  { \setlength{\itemsep}{0pt}
     \setlength{\parsep}{0pt}
    \setlength{\topsep}{0pt}
    \setlength{\partopsep}{0pt}
\setlength{\leftmargin}{2em}
\setlength{\labelwidth}{1.5em}
\setlength{\labelsep}{0.5em} } }

\newcommand{\squishend}{
\end{list}  }
\begin{document}

\title{Balancing Fine-tuning and RAG: A Hybrid Strategy for Dynamic LLM Recommendation Updates}

\author{Changping Meng}
\authornote{Equal Contribution}
\email{changping@gmail.com}
\affiliation{\institution{Google} \city{Mountain View} \state{California} \country{USA}}
\author{Hongyi Ling}
\authornotemark[1]
\email{linghongyi@google.com}
\affiliation{\institution{Google} \city{Mountain View} \state{California} \country{USA}}
\author{Jianling Wang}
\authornotemark[1]
\email{jianlingw@google.com}
\affiliation{\institution{Google Deepmind} \city{Mountain View} \state{California} \country{USA}}
\author{Yifan Liu}
\email{yifanliu@google.com}
\affiliation{\institution{Google} \city{Mountain View} \state{California} \country{USA}}
\author{Shuzhou Zhang}
\email{shuzhouz@google.com}
\affiliation{\institution{Google} \city{Mountain View} \state{California} \country{USA}}
\author{Dapeng Hong}
\email{dapengh@google.com}
\affiliation{\institution{Google} \city{Mountain View} \state{California} \country{USA}}
\author{Mingyan Gao}
\email{mingyan@google.com}
\affiliation{\institution{Google} \city{Mountain View} \state{California} \country{USA}}
\author{Onkar Dalal}
\email{onkardalal@google.com}
\affiliation{\institution{Google} \city{Mountain View} \state{California} \country{USA}}
\author{Ed Chi}
\email{edchi@google.com}
\affiliation{\institution{Google Deepmind} \city{Mountain View} \state{California} \country{USA}}
\author{Lichan Hong}
\email{lichan@google.com}
\affiliation{\institution{Google Deepmind} \city{Mountain View} \state{California} \country{USA}}
\author{Haokai Lu}
\email{haokai@google.com}
\affiliation{\institution{Google Deepmind} \city{Mountain View} \state{California} \country{USA}}
\author{Ningren Han}
\email{peterhan@google.com}
\affiliation{\institution{Google} \city{Mountain View} \state{California} \country{USA}}

\settopmatter{authorsperrow=4}

\renewcommand{\shortauthors}{Changping Meng et al.}

\begin{CCSXML}
<ccs2012>
<concept>
<concept_id>10002951.10003317</concept_id>
<concept_desc>Information systems~Information retrieval</concept_desc>
<concept_significance>500</concept_significance>
</concept>
<concept>
<concept_id>10010147.10010178</concept_id>
<concept_desc>Computing methodologies~Artificial intelligence</concept_desc>
<concept_significance>300</concept_significance>
</concept>
</ccs2012>
\end{CCSXML}

\ccsdesc[500]{Information systems~Information retrieval}
\ccsdesc[300]{Computing methodologies~Artificial intelligence}

\begin{abstract}
Large Language Models (LLMs) empower recommendation systems through their advanced reasoning and planning capabilities. However, the dynamic nature of user interests and content poses a significant challenge: 
While initial fine-tuning aligns LLMs with domain knowledge and user preferences, it fails to capture such real-time changes, necessitating robust update mechanisms. This paper investigates strategies for updating LLM-powered recommenders, focusing on the trade-offs between ongoing fine-tuning and Retrieval-Augmented Generation (RAG). Using an LLM-powered user interest exploration system as a case study, we perform a comparative analysis of these methods across dimensions like cost, agility, and knowledge incorporation. 
We propose a hybrid update strategy that leverages the long-term knowledge adaptation of periodic fine-tuning with the agility of low-cost RAG. We demonstrate through live A/B experiments on a billion-user platform that this hybrid approach yields statistically significant improvements in user satisfaction, offering a practical and cost-effective framework for maintaining high-quality LLM-powered recommender systems.
\end{abstract}

\keywords{Large Language Models, Recommendation System, User Interest Exploration}

\maketitle
% \vspace{-0.1in}

\section{Introduction}

The emergence of Large Language Models (LLMs) is transforming the landscape of recommendation systems with their extensive world knowledge and reasoning capabilities. LLM-powered recommenders~\cite{kim2024large,acharya2023llm,xu2024openp5} utilize the deep semantic understanding and generative strengths of these models to deliver more personalized, explainable, and context-aware suggestions. 

A common approach involves initially fine-tuning an LLM on domain knowledge and historical user-item interactions to tailor it for specific recommendation tasks \cite{wang2024llms}. However, the environments where these systems operate are inherently dynamic \cite{wang2020next,wang2020time}. User interests evolve, new items emerge constantly, and underlying data patterns shift – for instance, analysis of user transitions between interest clusters often reveals significant temporal variability. An LLM fine-tuned only on past data captures a static snapshot and cannot inherently reflect these real-time dynamics. 

To address this challenge, two prominent techniques for adapting and updating LLMs are fine-tuning~\cite{LLMvie,chen2025dlcrec} and Retrieval-Augmented Generation (RAG)~\cite{zeng2024federated, fan2024right}. Fine-tuning involves further training a pre-trained LLM on a specific dataset to adjust its internal parameters, tailoring its knowledge or behavior. RAG, conversely, connects the LLM to external knowledge sources at inference time, retrieving relevant information to augment the prompt and ground the model's generation in specific, often up-to-date, data without altering the model's parameters.

This paper conducts a comparative analysis of fine-tuning and RAG as methodologies for adapting LLM-powered recommendation systems to dynamic updates. Our investigation is grounded in a deployed LLM-powered user interest recommendation system~\cite{LLMvie, wang2024llms}. While interest exploration systems~\cite{chen2021values, chen2021exploration, song2022show, mahajan2023pie, su2024long} aim to diversify recommendations, effectively introducing novel interests poses a significant challenge. In our case study, the fine-tuned LLM generates potential novel interest clusters from user history; the core update challenge we address is enabling this model to accurately reflect the changing popularity and relationships between these clusters over time.

This challenge leads to our central hypothesis: In a highly dynamic domain like short-form video recommendation, a static, fine-tuned LLM is insufficient to maintain recommendation quality over time. We hypothesize that a hybrid strategy, combining periodic fine-tuning with frequent RAG-based updates, will more effectively adapt to shifting user interest patterns and result in superior online performance. This paper tests this hypothesis through a LLM-powered user interest exploration system~\cite{LLMvie}. We therefore compare fine-tuning and RAG specifically for this task, discussing their respective system designs, processes, strengths, limitations, effectiveness, and cost, using both offline and live experimental results.

% \vspace{-0.1in}
\section{Method}
This section first provides necessary preliminary information and outlines the motivation for our work, followed by a detailed description of the interest exploration system. Subsequently, we detail the designs for fine-tuning and RAG.

\subsection{Preliminary}

\noindent\textbf{Motivation.} To effectively model the dynamic nature of user interests, we represent them using clusters, following the methodology in \cite{chang2024cluster}. 
To assess the evolution of user interest transitions, we first define a `successor interest'. From user interaction logs, we construct sequences of three consecutive, distinct item clusters a user engages with, denoted as 
$(c1, c2, c_{next})$. Here, $c_{next}$ is the `successor interest' to the preceding pair $(c1, c2)$. We then measured the stability of the top-5 most frequent successor interests month-over-month using the Jaccard Similarity (i.e., quantifying the semantic overlap between these top-5 sets). 
Our analysis revealed substantial variability, with a low mean Jaccard score of 0.17 (variance 0.07), demonstrating substantial monthly variability in prevalent user transition patterns. This observed dynamism highlights the critical need for efficiently incorporating refreshed user feedback.

\smallskip
\noindent\textbf{Interest Exploration System.} 
In the LLM-powered system~\cite{LLMvie}, each user's recent interaction history is represented as a sequence of $k$ interest clusters $S_u = \{c_1, c_2, \dots, c_k\}$, where each $c_i \in \mathcal{C}$ denotes an item interest cluster from a predefined cluster set $\mathcal{C}$ \cite{chang2024cluster}. Each interest cluster groups items that are topically coherent, based on their metadata and content feature. Given $S_u$, the LLM predicts the user’s next novel interest cluster $c_n \in \mathcal{C}$. 
Because online serving the LLM for a billion-user system is prohibitively costly, we precompute and store the predicted next-cluster transitions for all possible $k$-length sequences of interest clusters. Let $\mathcal{S} = \{(c_1, \dots, c_k) \mid c_i \in \mathcal{C}\}$ denote the set of all possible $k$-length cluster sequences. For each $S \in \mathcal{S}$, we store a corresponding predicted novel cluster $c_n$ offline. During online serving, a user’s current history $S_u$ is matched to a set $S \in \mathcal{S}$, and the corresponding predicted next cluster is retrieved via table lookup.

\subsection{Fine-tune}

Following the preliminary example~\cite{LLMvie}, with $k=2$, each fine-tuning data sample is denoted as $[(c_1,c_2 ), c_{next}]$. The prompt is illustrated as black lines in Figure~\ref{fig:prompt}. Periodically, we curate thousands of those pairs for fine-tuning. Fine-tuning offers the benefits of adapting model behavior and style, as well as improving performance on specific tasks. However, the drawbacks are also significant, including high cost and complexity, and the risk of overfitting. Due to the high cost of fine-tuning, updates happen on a monthly basis.

We also propose two key evaluation metrics to evaluate the fine-tuning quality: the exact match rate (percentage of predictions precisely matching the partition description) and the test set recall (percentage of predictions aligning with users' watch history). Leveraging these insights, our auto-refreshed fine-tuning pipeline implements two automated quality checks:
\squishlist
    \item If the exact match rate during partition mapping generation is below 90\%, the pipeline execution is halted.
    \item If the test set recall is less than 1.5\%, the pipeline fails.
\squishend
These conditions necessitate manual review by an engineer to identify the root cause and decide whether to proceed to production or re-run the process.

\subsection{RAG}
\begin{figure}
    % \vspace{-0.1in}
\centering
    \includegraphics[width=.45\textwidth]{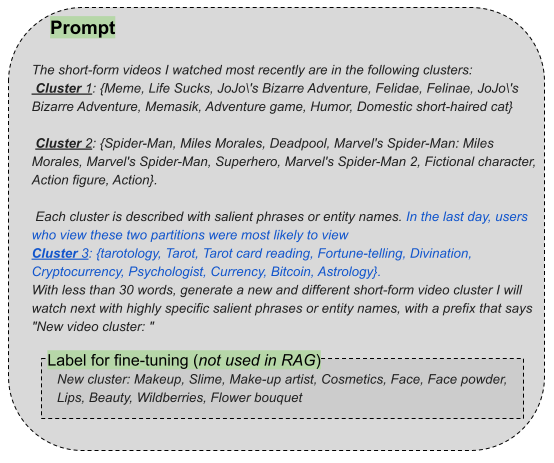}
    \vspace{-0.1in}
    \caption{ Prompt for Novel Interest Prediction. \textbf{Black} lines are the fine-tuning prompt. Added \textcolor{blue}{\textbf{Blue}} lines are the RAG version with injected recent watch history. Label is only used for fine-tuning, but not RAG}
    \vspace{-0.1in}
\label{fig:prompt}
\end{figure}

\begin{figure}[h]
\vspace{-0.1in}
\centering
    \includegraphics[width=.45\textwidth]{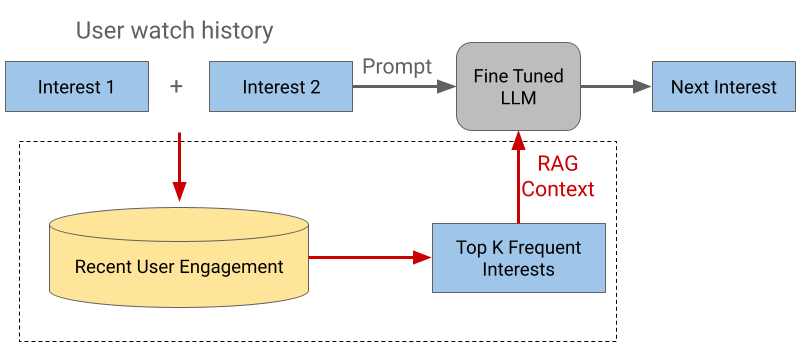}
    \vspace{-0.2in}
    \caption{Refresh with Retrieval-Augmented Generation.  Fine-tuning is refreshed monthly, while RAG is refreshed multiple times within a week.}
\label{fig:flow}
\vspace{-0.2in}
\end{figure}

Instead of retraining LLM with new viewing data with high cost, we can prompt new data to the LLM and perform bulk inference periodically to generate a dynamic transition mapping with low cost. Adhering to the prompt design of the LLM-powered interest exploration system~\cite{LLMvie}, we represent a user's consumption history as a sequence of their most recently interacted unique clusters. Each cluster is defined by a set of keywords. To better capture both dynamic system-wide trends and individual user's evolving preferences, these prompts incorporate top popular interest clusters along with the user's recent watch history, as detailed in Figure~\ref{fig:prompt}. Fine-tuning can be done on a monthly schedule, while the RAG prompt can happen more frequently, even at daily basis, with the overall system illustrated in Figure~\ref{fig:flow}.

% https://docs.google.com/drawings/d/1BPHHGa3EhciqGSCPITruKZ0GAxsePcLZ7pSXWmVkUo8/edit?usp=sharing

\subsubsection{Granularity}

During the bulk inference phase, RAG prompts can be generated at different level of granularity. 
\squishlist
    \item Instance level. Prompts are tailored for each individual cluster pair. Specifically, we can identify the top-1 most frequent next cluster based on recent data. Consider a distribution $\{(c_1, c_2, c_3): 10, (c_1, c_2, c_4): 8, (c_1, c_2, c_5): 2, \dots\}$. Since $c_3$ appears most frequently following $c_1$ and $c_2$, $c_3$ can be included in the prompt for inference.
    
    \item Global level. This approach uses a single, universal prompt for all data pairs. This prompt captures overall user behavior and might include illustrative examples. 
    E.g. we can construct the prompt using the top-100 most frequent pairs found across the entire new data, regardless of specific input pairs. These globally representative clusters are then included to guide inference.
\squishend

Given that the global level design might introduce noise to the target cluster pair during output cluster generation, we adopt the instance level design approach. 

\subsubsection{Retrieval Similarity}

This section outlines methods for retrieving relevant recent data during bulk inference.

\squishlist
 \item Frequency-based Retrieval: We identify data points within the same cluster pair as the query and select those with the highest frequency. This provides the LLM with prompts reflecting recent, prevalent user behaviors for the specific cluster.
 \item Trend-based Retrieval: Focusing on the query's cluster pair, we select data points exhibiting the largest frequency difference, highlighting emerging or declining user interests.
\squishend
Our analysis and evaluation indicate that frequency-based retrieval yields the best results.

Number of retrieved clusters included in the context can vary (e.g. the Cluster 3 in Figure~\ref{fig:prompt}). While a larger number provides richer information, it also increases computational cost. Our live experiments suggest that including the top 1 most frequent cluster is sufficient to provide satisfying results.

\subsection{Data Retrieval}
We use users' interaction history, represented as a sequence of watches, on a large-scale video platform as the source dataset. Our data extraction targets videos demonstrating positive viewer engagement. 
To refine the dataset, we deduplicate video cluster IDs within each sequence and remove sequences with fewer than two videos.
For the remaining sequences, we construct tuples of three consecutive video cluster IDs as $(c_1, c_2, c_{next})$. The final step is to count the occurrences of the next cluster for each cluster pair. 

\section{Results and Evaluation}
In our hybrid update strategy, LLM models undergo monthly fine-tuning, while RAG refresh occurs sub-weekly. From the fine-tuned model, we then measure the incremental gains of more frequent and up-to-date RAG.
\subsection{Offline Evaluation}

\begin{figure}[htbp]
\vspace{-0.2in}
    \centering
    \begin{subfigure}[t]{0.47\linewidth}   
        \centering
        \includegraphics[width=\linewidth]{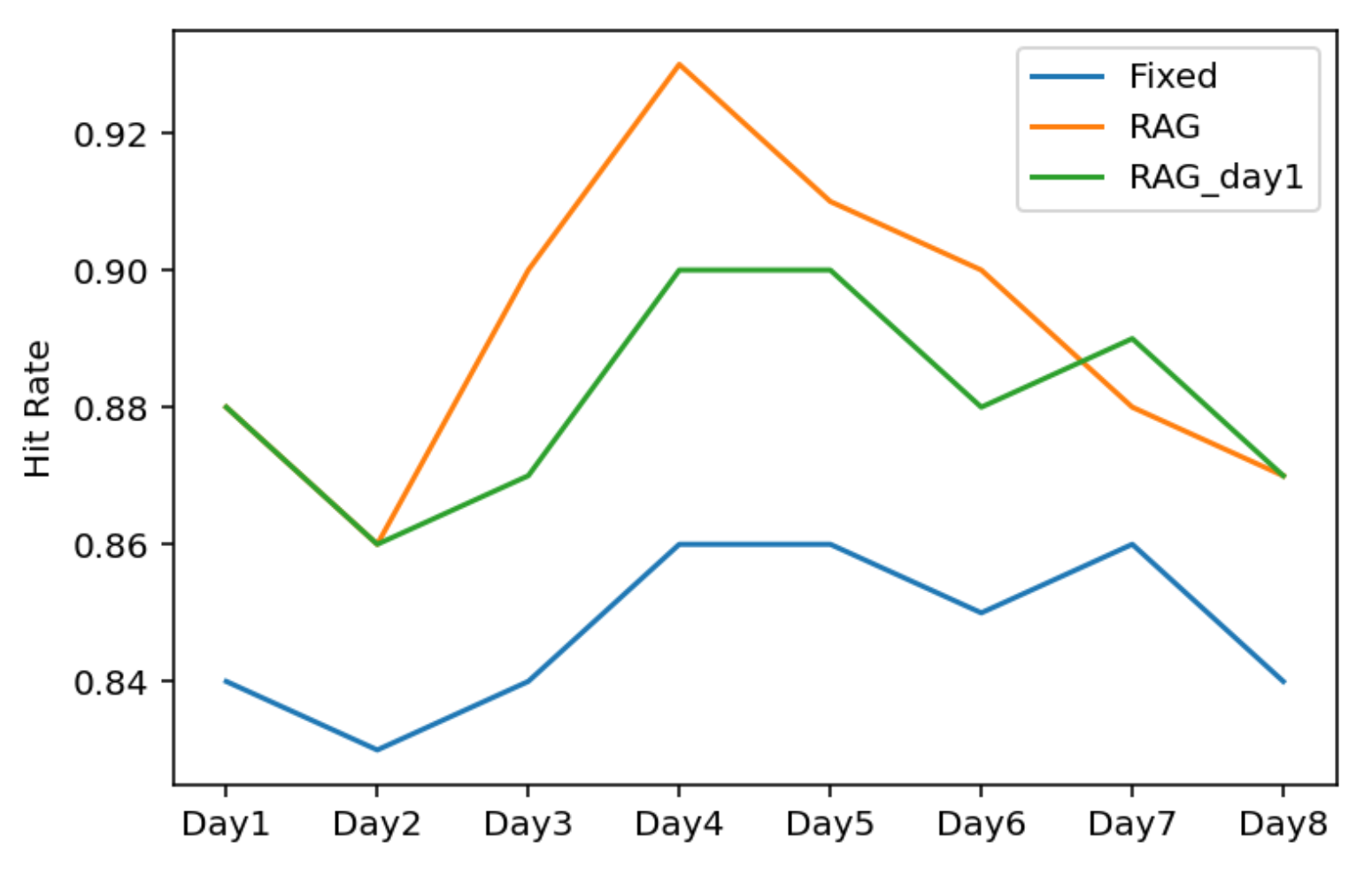}
        \vspace{-0.2in}
        \caption{Trajectory of hit rate.}
        \label{fig:hit}
    \end{subfigure}
    \quad
    \begin{subfigure}[t]{0.47\linewidth}   
        \centering
        \includegraphics[width=\linewidth]{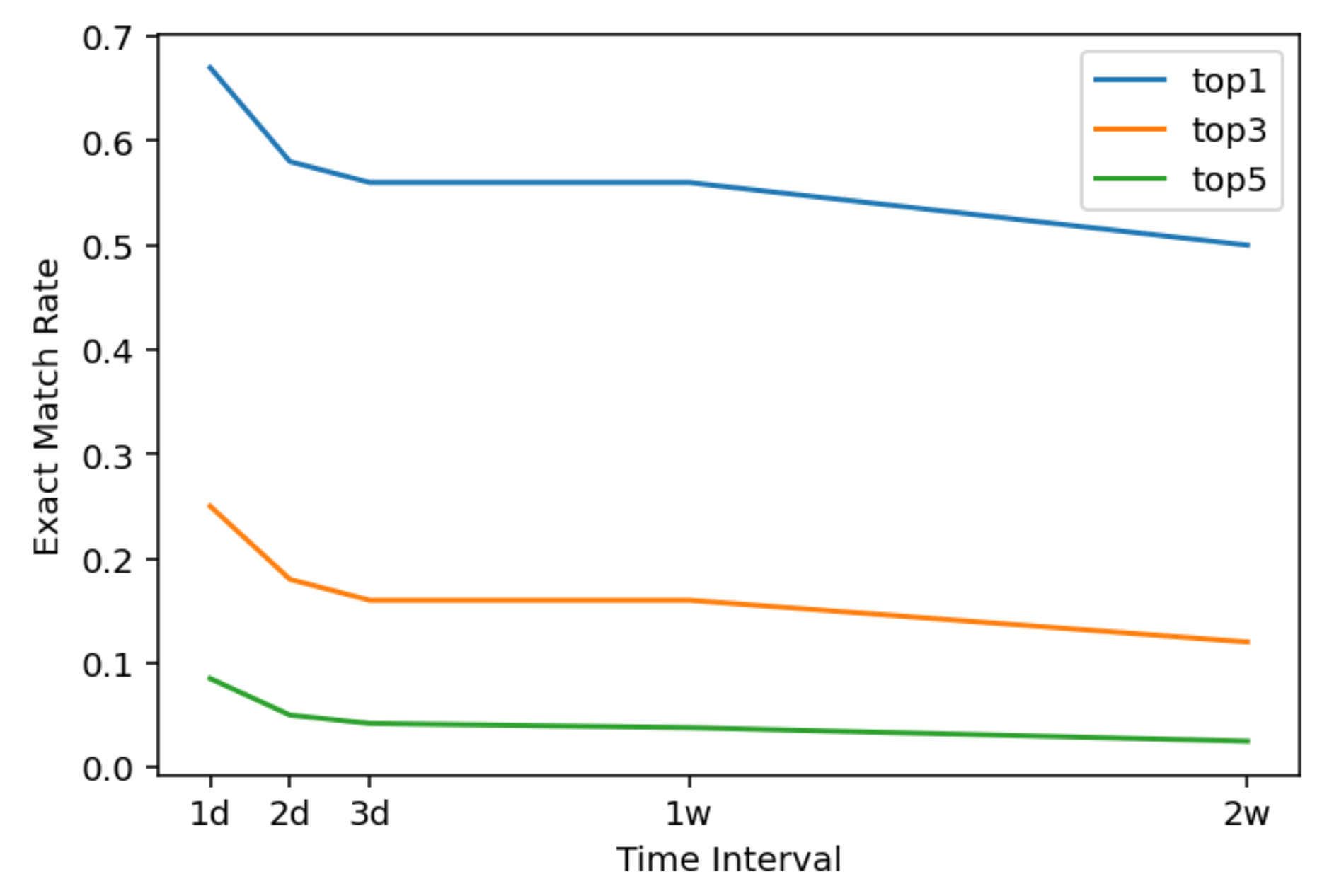}
        \vspace{-0.2in}
        \caption{Exact match rates for top k clusters over time}
        \label{fig:identity_rate}
    \end{subfigure}
    \vspace{-0.1 in}
    \caption{Offline Evaluation.}
    \vspace{-0.2 in}
    \label{fig:offline_eval}
\end{figure}

We evaluated how RAG-generated cluster mappings evolve over time and their alignment with user behavior. Specifically, We assessed the hit rate, which computes the proportion of times the predicted next cluster appears in the real user sequence. We compared three versions of transition mappings: (1) a fixed mapping generated without RAG; (2) a RAG-generated mapping updated every two days; and (3) a RAG-generated mapping computed only on $day_1$ and held fixed thereafter. As illustrated in Figure~\ref{fig:hit}, both RAG-based mappings outperform the fixed baseline, with the version updated every two days achieving slightly higher hit rates.

To better understand the influence of RAG on the LLM’s generation behavior, we analyze the similarity between outputs generated with and without RAG. Only 7.8\% of the RAG-generated outputs were identical to those produced without RAG, compared to a 37.5\% overlap when using repeated prompts without RAG. The results indicate that RAG significantly alters the generated content, often leading to novel predictions that differ from both the retrieved context and the non-RAG outputs.

Finally, we studied how the top-$k$ most frequent clusters for each cluster pair changed over time. Our findings reveal a significant shift in top clusters across retrieval dates, with substantial drops in overlap as time progresses. This trend, illustrated in Figure~\ref{fig:identity_rate}, re-emphasize the dynamic nature of user interests and underscores the need for regularly refreshed retrieval to reflect current behavioral patterns.

\vspace{-0.2in}
\subsection{Live Experiment}

We conducted A/B experiments within a short-form video recommendation system, serving billions of users, to measure the effectiveness of RAG in enhancing the performance of our LLM-powered interest exploration system. Gemini 1.5~\cite{team2024gemini} was adopted as the base LLM for this system, while the process and pipeline are designed for adaptability to other models. The system's high-level function recommends novel interest clusters, currently based on a user's historical interest cluster sequence of length $K=2$.

\begin{figure}[htbp]
    \centering
    \begin{subfigure}[b]{0.46\linewidth}   
        \includegraphics[width=\linewidth]{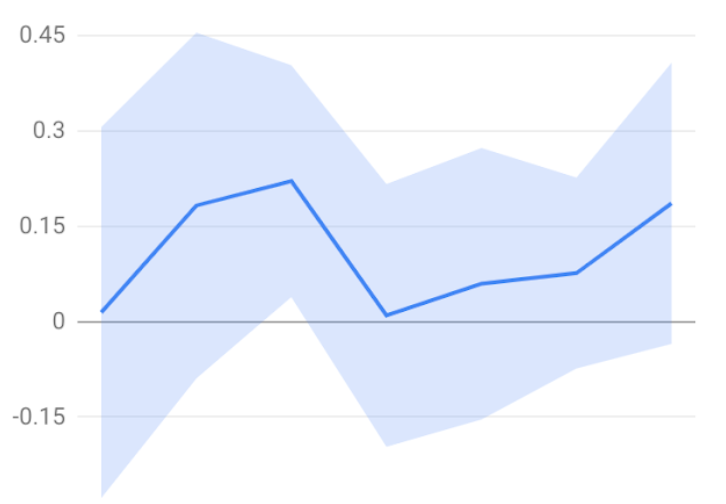}
        \vspace{-0.3in}
        \caption{Satisfied User Outcomes}
        \label{fig:seu}
    \end{subfigure}
    \quad
    \begin{subfigure}[b]{0.46\linewidth}   
        \centering
        \includegraphics[width=\linewidth]{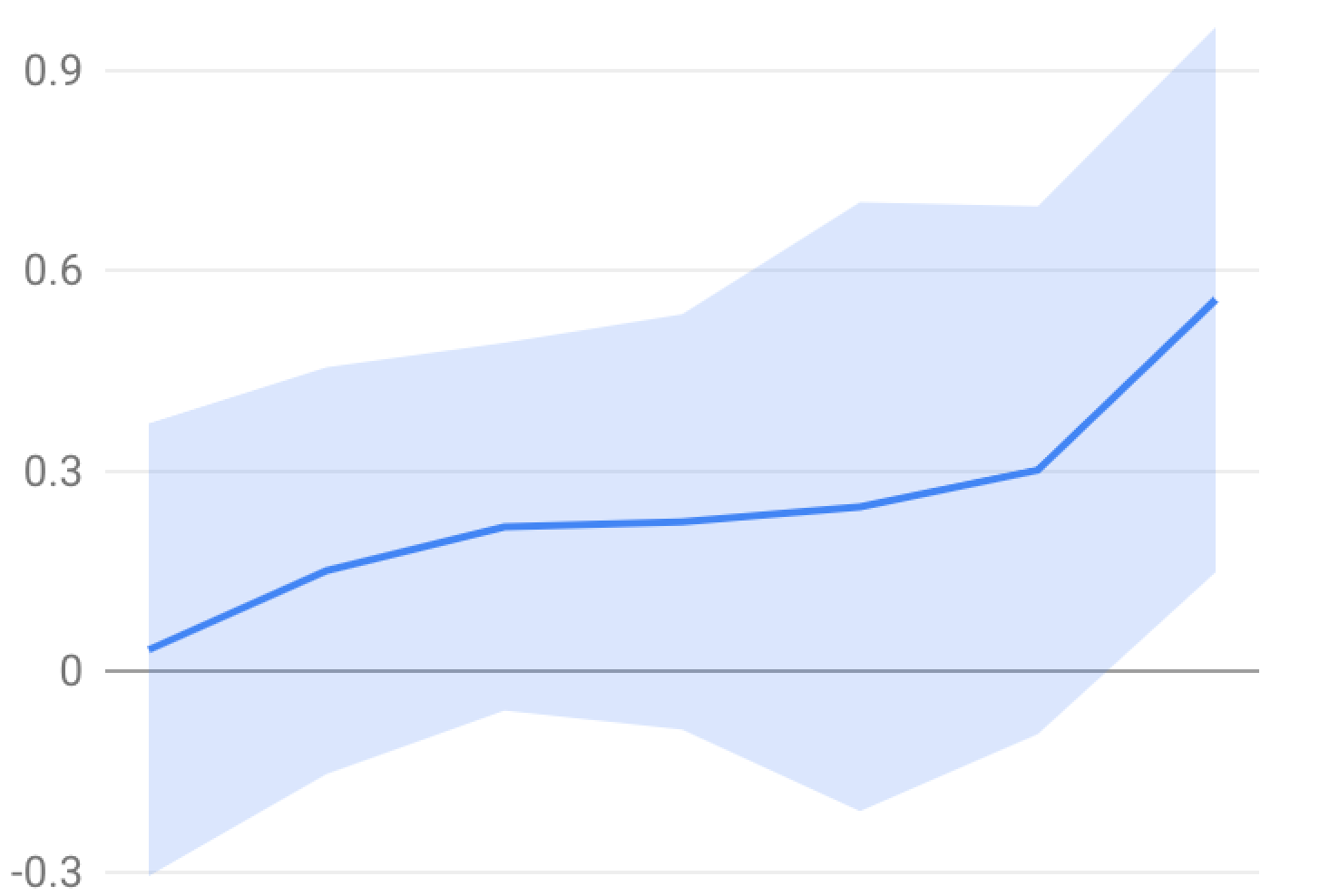}
        \vspace{-0.3in}
        \caption{Satisfaction Rate}
        \label{fig:like}
    \end{subfigure}
%   \vspace{-0.1in} % Add some vertical space here
    \\
    \begin{subfigure}[b]{0.46\linewidth}   
        \includegraphics[width=\linewidth]{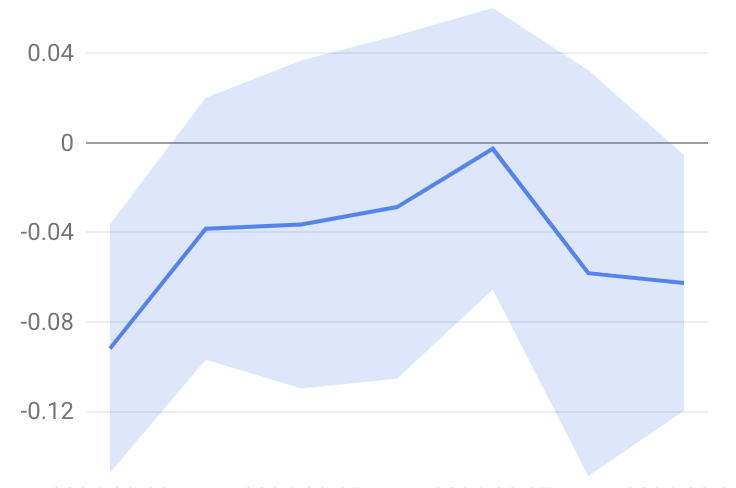}
        \vspace{-0.3in}
        \caption{Dissatisfaction Rate}
        \label{fig:skip}
    \end{subfigure}
    \quad
    \begin{subfigure}[b]{0.46\linewidth}   
        \centering
        \includegraphics[width=\linewidth]{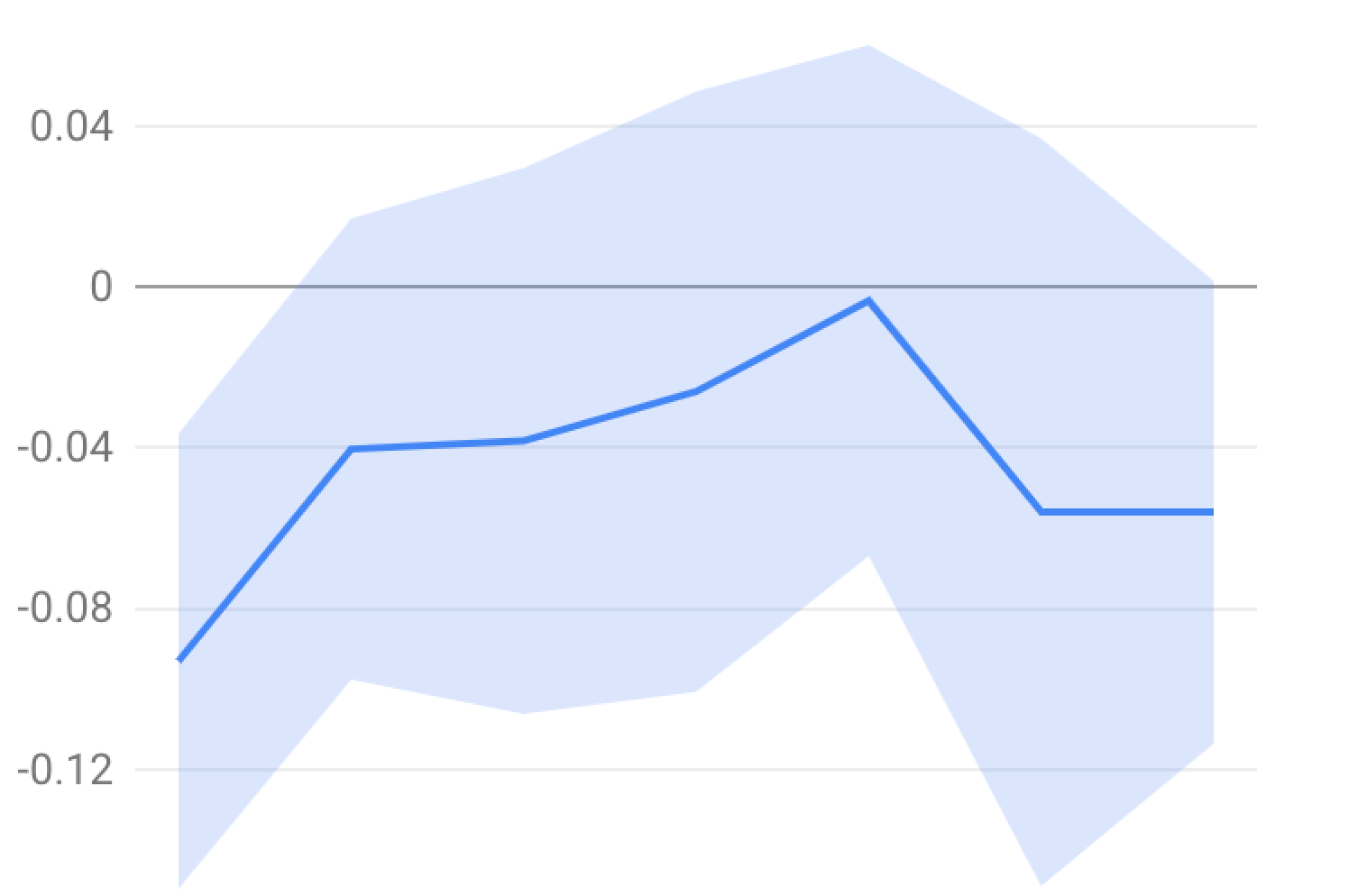}
        \vspace{-0.3in}
        \caption{Negative Interaction}
        \label{fig:negative}
    \end{subfigure}
    \vspace{-0.1in}
    \caption{Live experiment results for user metrics. The x-axis represents the date; the y-axis represents the relative difference (in percentage) between the treatment and control groups.}
    \label{fig:live_exp}
\end{figure}

We report the user metrics of the live experiments in Figure~\ref{fig:live_exp}. The x-axis represents the date, and the y-axis shows the relative percentage difference between the treatment and control. We also report the mean and 95\% confidence intervals for each metric. The top-tier metrics \textit{Satisfied User Outcomes} are increased by $0.11\%$ with 95\% confidence interval $[0.00\%, 0.21\%]$, which is highly significant at the scale of our system.  \textit{Satisfaction Rate} is increased by $0.25\%$ with interval $[0.01\%, 0.48\%]$. The \textit{Dissatisfaction Rate} is reduced by $0.05\%$ with interval $[-0.08\%, -0.01\%]$. \textit{Negative Interaction} is reduced by $0.04\%$ with interval $[-0.08\%, -0.01\%]$.

 We employed RAG to update the cluster transition table on $day_1$ and $day_4$. Following these updates, we observed notable increases in user engagement, including significant improvements in Satisfied User Outcomes and Satisfaction Rate, indicating enhanced user satisfaction.

% Rasta: http://exp/analysis/_:FcQe5VUldtTc-NKQ6bx2hR91dew

\vspace{-0.1in}

\section{Conclusion}
This paper investigated the critical challenge of keeping LLM-powered recommendation systems updated. We conducted a comparative analysis of fine-tuning and RAG, proposing and validating a hybrid strategy. Our core finding is that combining monthly fine-tuning with sub-weekly RAG updates provides a robust, cost-effective solution for adapting to dynamic user interests, leading to significant improvements in online user satisfaction metrics in a large-scale production environment. Future work will explore more adaptive update cadences, where the frequency of RAG or fine-tuning is determined automatically based on the detected rate of interest drift. 

\vspace{-0.05in}
\section*{Speaker Bio}
Changping Meng is a software engineer at Google (YouTube). He received the Computer Science PhD from the Purdue University. His work primarily focuses on short-form video recommendations.
\vspace{-0.2in}

\bibliographystyle{ACM-Reference-Format} % We choose the "plain" reference style
\bibliography{refs} % Entries are in the refs.bib file


\begin{thebibliography}{17}

%%% the Web address.

\ifx \showCODEN    \undefined \def \showCODEN     #1{\unskip}     \fi
\ifx \showDOI      \undefined \def \showDOI       #1{#1}\fi
\ifx \showISBNx    \undefined \def \showISBNx     #1{\unskip}     \fi
\ifx \showISBNxiii \undefined \def \showISBNxiii  #1{\unskip}     \fi
\ifx \showISSN     \undefined \def \showISSN      #1{\unskip}     \fi
\ifx \showLCCN     \undefined \def \showLCCN      #1{\unskip}     \fi
\ifx \shownote     \undefined \def \shownote      #1{#1}          \fi
\ifx \showarticletitle \undefined \def \showarticletitle #1{#1}   \fi
\ifx \showURL      \undefined \def \showURL       {\relax}        \fi
% invisible to TeX
\providecommand\bibfield[2]{#2}
\providecommand\bibinfo[2]{#2}
\providecommand\natexlab[1]{#1}
\providecommand\showeprint[2][]{arXiv:#2}

\bibitem[\protect\citeauthoryear{Acharya, Singh, and Onoe}{Acharya
  et~al\mbox{.}}{2023}]%
        {acharya2023llm}
\bibfield{author}{\bibinfo{person}{Arkadeep Acharya}, \bibinfo{person}{Brijraj
  Singh}, {and} \bibinfo{person}{Naoyuki Onoe}.}
  \bibinfo{year}{2023}\natexlab{}.
\newblock \showarticletitle{Llm based generation of item-description for
  recommendation system}. In \bibinfo{booktitle}{\emph{Proceedings of the 17th
  ACM conference on recommender systems}}. \bibinfo{pages}{1204--1207}.
\newblock


\bibitem[\protect\citeauthoryear{Chang, Meng, Ma, Chang, Gu, Peng, Feng, Zhang,
  Bi, Chi, and Chen}{Chang et~al\mbox{.}}{2024}]%
        {chang2024cluster}
\bibfield{author}{\bibinfo{person}{Bo Chang}, \bibinfo{person}{Changping Meng},
  \bibinfo{person}{He Ma}, \bibinfo{person}{Shuo Chang}, \bibinfo{person}{Yang
  Gu}, \bibinfo{person}{Yajun Peng}, \bibinfo{person}{Jingchen Feng},
  \bibinfo{person}{Yaping Zhang}, \bibinfo{person}{Shuchao Bi},
  \bibinfo{person}{Ed~H Chi}, {and} \bibinfo{person}{Minmin Chen}.}
  \bibinfo{year}{2024}\natexlab{}.
\newblock \showarticletitle{Cluster Anchor Regularization to Alleviate
  Popularity Bias in Recommender Systems}. In
  \bibinfo{booktitle}{\emph{Companion Proceedings of the ACM Web Conference
  2024}}.
\newblock


\bibitem[\protect\citeauthoryear{Chen, Gao, Yuan, Liu, Cai, and Jiang}{Chen
  et~al\mbox{.}}{2025}]%
        {chen2025dlcrec}
\bibfield{author}{\bibinfo{person}{Jiaju Chen}, \bibinfo{person}{Chongming
  Gao}, \bibinfo{person}{Shuai Yuan}, \bibinfo{person}{Shuchang Liu},
  \bibinfo{person}{Qingpeng Cai}, {and} \bibinfo{person}{Peng Jiang}.}
  \bibinfo{year}{2025}\natexlab{}.
\newblock \showarticletitle{DLCRec: A Novel Approach for Managing Diversity in
  LLM-Based Recommender Systems}. In \bibinfo{booktitle}{\emph{Proceedings of
  the Eighteenth ACM International Conference on Web Search and Data Mining}}.
  \bibinfo{pages}{857--865}.
\newblock


\bibitem[\protect\citeauthoryear{Chen}{Chen}{2021}]%
        {chen2021exploration}
\bibfield{author}{\bibinfo{person}{Minmin Chen}.}
  \bibinfo{year}{2021}\natexlab{}.
\newblock \showarticletitle{Exploration in recommender systems}. In
  \bibinfo{booktitle}{\emph{Proceedings of the 15th ACM Conference on
  Recommender Systems}}. \bibinfo{pages}{551--553}.
\newblock


\bibitem[\protect\citeauthoryear{Chen, Wang, Xu, Le, Sharma, Richardson, Wu,
  and Chi}{Chen et~al\mbox{.}}{2021}]%
        {chen2021values}
\bibfield{author}{\bibinfo{person}{Minmin Chen}, \bibinfo{person}{Yuyan Wang},
  \bibinfo{person}{Can Xu}, \bibinfo{person}{Ya Le}, \bibinfo{person}{Mohit
  Sharma}, \bibinfo{person}{Lee Richardson}, \bibinfo{person}{Su-Lin Wu}, {and}
  \bibinfo{person}{Ed Chi}.} \bibinfo{year}{2021}\natexlab{}.
\newblock \showarticletitle{Values of user exploration in recommender systems}.
  In \bibinfo{booktitle}{\emph{Proceedings of the 15th acm Conference on
  recommender systems}}. \bibinfo{pages}{85--95}.
\newblock


\bibitem[\protect\citeauthoryear{Fan, Fan, Chen, Guo, Zhang, and Cheng}{Fan
  et~al\mbox{.}}{2024}]%
        {fan2024right}
\bibfield{author}{\bibinfo{person}{Run-Ze Fan}, \bibinfo{person}{Yixing Fan},
  \bibinfo{person}{Jiangui Chen}, \bibinfo{person}{Jiafeng Guo},
  \bibinfo{person}{Ruqing Zhang}, {and} \bibinfo{person}{Xueqi Cheng}.}
  \bibinfo{year}{2024}\natexlab{}.
\newblock \showarticletitle{RIGHT: Retrieval-augmented generation for
  mainstream hashtag recommendation}. In \bibinfo{booktitle}{\emph{European
  Conference on Information Retrieval}}. Springer, \bibinfo{pages}{39--55}.
\newblock


\bibitem[\protect\citeauthoryear{Kim, Kang, Choi, Kim, Yang, and Park}{Kim
  et~al\mbox{.}}{2024}]%
        {kim2024large}
\bibfield{author}{\bibinfo{person}{Sein Kim}, \bibinfo{person}{Hongseok Kang},
  \bibinfo{person}{Seungyoon Choi}, \bibinfo{person}{Donghyun Kim},
  \bibinfo{person}{Minchul Yang}, {and} \bibinfo{person}{Chanyoung Park}.}
  \bibinfo{year}{2024}\natexlab{}.
\newblock \showarticletitle{Large language models meet collaborative filtering:
  An efficient all-round llm-based recommender system}. In
  \bibinfo{booktitle}{\emph{Proceedings of the 30th ACM SIGKDD Conference on
  Knowledge Discovery and Data Mining}}. \bibinfo{pages}{1395--1406}.
\newblock


\bibitem[\protect\citeauthoryear{Mahajan, Porobo~Dharwadker, Shah, Qu, Bang,
  and Schumitsch}{Mahajan et~al\mbox{.}}{2023}]%
        {mahajan2023pie}
\bibfield{author}{\bibinfo{person}{Khushhall~Chandra Mahajan},
  \bibinfo{person}{Amey Porobo~Dharwadker}, \bibinfo{person}{Romil Shah},
  \bibinfo{person}{Simeng Qu}, \bibinfo{person}{Gaurav Bang}, {and}
  \bibinfo{person}{Brad Schumitsch}.} \bibinfo{year}{2023}\natexlab{}.
\newblock \showarticletitle{PIE: Personalized Interest Exploration for
  Large-Scale Recommender Systems}. In \bibinfo{booktitle}{\emph{Companion
  Proceedings of the ACM Web Conference 2023}}. \bibinfo{pages}{508--512}.
\newblock


\bibitem[\protect\citeauthoryear{Song, Sun, Lian, Huang, Li, Jin, and Xie}{Song
  et~al\mbox{.}}{2022}]%
        {song2022show}
\bibfield{author}{\bibinfo{person}{Yu Song}, \bibinfo{person}{Shuai Sun},
  \bibinfo{person}{Jianxun Lian}, \bibinfo{person}{Hong Huang},
  \bibinfo{person}{Yu Li}, \bibinfo{person}{Hai Jin}, {and}
  \bibinfo{person}{Xing Xie}.} \bibinfo{year}{2022}\natexlab{}.
\newblock \showarticletitle{Show me the whole world: Towards entire item space
  exploration for interactive personalized recommendations}. In
  \bibinfo{booktitle}{\emph{Proceedings of the Fifteenth ACM International
  Conference on Web Search and Data Mining}}. \bibinfo{pages}{947--956}.
\newblock


\bibitem[\protect\citeauthoryear{Su, Wang, Le, Liu, Li, Lu, Lipshitz, Badam,
  Heldt, Bi, et~al\mbox{.}}{Su et~al\mbox{.}}{2024}]%
        {su2024long}
\bibfield{author}{\bibinfo{person}{Yi Su}, \bibinfo{person}{Xiangyu Wang},
  \bibinfo{person}{Elaine~Ya Le}, \bibinfo{person}{Liang Liu},
  \bibinfo{person}{Yuening Li}, \bibinfo{person}{Haokai Lu},
  \bibinfo{person}{Benjamin Lipshitz}, \bibinfo{person}{Sriraj Badam},
  \bibinfo{person}{Lukasz Heldt}, \bibinfo{person}{Shuchao Bi},
  {et~al\mbox{.}}} \bibinfo{year}{2024}\natexlab{}.
\newblock \showarticletitle{Long-term value of exploration: Measurements,
  findings and algorithms}. In \bibinfo{booktitle}{\emph{Proceedings of the
  17th ACM International Conference on Web Search and Data Mining}}.
  \bibinfo{pages}{636--644}.
\newblock


\bibitem[\protect\citeauthoryear{Team, Georgiev, Lei, Burnell, Bai, Gulati,
  Tanzer, Vincent, Pan, Wang, et~al\mbox{.}}{Team et~al\mbox{.}}{2024}]%
        {team2024gemini}
\bibfield{author}{\bibinfo{person}{Gemini Team}, \bibinfo{person}{Petko
  Georgiev}, \bibinfo{person}{Ving~Ian Lei}, \bibinfo{person}{Ryan Burnell},
  \bibinfo{person}{Libin Bai}, \bibinfo{person}{Anmol Gulati},
  \bibinfo{person}{Garrett Tanzer}, \bibinfo{person}{Damien Vincent},
  \bibinfo{person}{Zhufeng Pan}, \bibinfo{person}{Shibo Wang}, {et~al\mbox{.}}}
  \bibinfo{year}{2024}\natexlab{}.
\newblock \showarticletitle{Gemini 1.5: Unlocking multimodal understanding
  across millions of tokens of context}.
\newblock \bibinfo{journal}{\emph{arXiv preprint arXiv:2403.05530}}
  (\bibinfo{year}{2024}).
\newblock


\bibitem[\protect\citeauthoryear{Wang, Ding, Hong, Liu, and Caverlee}{Wang
  et~al\mbox{.}}{2020a}]%
        {wang2020next}
\bibfield{author}{\bibinfo{person}{Jianling Wang}, \bibinfo{person}{Kaize
  Ding}, \bibinfo{person}{Liangjie Hong}, \bibinfo{person}{Huan Liu}, {and}
  \bibinfo{person}{James Caverlee}.} \bibinfo{year}{2020}\natexlab{a}.
\newblock \showarticletitle{Next-item recommendation with sequential
  hypergraphs}. In \bibinfo{booktitle}{\emph{Proceedings of the 43rd
  international ACM SIGIR conference on research and development in information
  retrieval}}. \bibinfo{pages}{1101--1110}.
\newblock


\bibitem[\protect\citeauthoryear{Wang, Louca, Hu, Cellier, Caverlee, and
  Hong}{Wang et~al\mbox{.}}{2020b}]%
        {wang2020time}
\bibfield{author}{\bibinfo{person}{Jianling Wang}, \bibinfo{person}{Raphael
  Louca}, \bibinfo{person}{Diane Hu}, \bibinfo{person}{Caitlin Cellier},
  \bibinfo{person}{James Caverlee}, {and} \bibinfo{person}{Liangjie Hong}.}
  \bibinfo{year}{2020}\natexlab{b}.
\newblock \showarticletitle{Time to shop for valentine's day: Shopping
  occasions and sequential recommendation in e-commerce}. In
  \bibinfo{booktitle}{\emph{Proceedings of the 13th international conference on
  web search and data mining}}. \bibinfo{pages}{645--653}.
\newblock


\bibitem[\protect\citeauthoryear{Wang, Lu, Liu, Ma, Wang, Gu, Zhang, Bi,
  Baugher, Chi, et~al\mbox{.}}{Wang et~al\mbox{.}}{2024a}]%
        {wang2024llms}
\bibfield{author}{\bibinfo{person}{Jianling Wang}, \bibinfo{person}{Haokai Lu},
  \bibinfo{person}{Yifan Liu}, \bibinfo{person}{He Ma}, \bibinfo{person}{Yueqi
  Wang}, \bibinfo{person}{Yang Gu}, \bibinfo{person}{Shuzhou Zhang},
  \bibinfo{person}{Shuchao Bi}, \bibinfo{person}{Lexi Baugher},
  \bibinfo{person}{Ed Chi}, {et~al\mbox{.}}} \bibinfo{year}{2024}\natexlab{a}.
\newblock \showarticletitle{LLMs for User Interest Exploration: A Hybrid
  Approach}.
\newblock \bibinfo{journal}{\emph{arXiv e-prints}} (\bibinfo{year}{2024}),
  \bibinfo{pages}{arXiv--2405}.
\newblock


\bibitem[\protect\citeauthoryear{Wang, Lu, Liu, Ma, Wang, Gu, Zhang, Han, Bi,
  Baugher, Chi, and Chen}{Wang et~al\mbox{.}}{2024b}]%
        {LLMvie}
\bibfield{author}{\bibinfo{person}{Jianling Wang}, \bibinfo{person}{Haokai Lu},
  \bibinfo{person}{Yifan Liu}, \bibinfo{person}{He Ma}, \bibinfo{person}{Yueqi
  Wang}, \bibinfo{person}{Yang Gu}, \bibinfo{person}{Shuzhou Zhang},
  \bibinfo{person}{Ningren Han}, \bibinfo{person}{Shuchao Bi},
  \bibinfo{person}{Lexi Baugher}, \bibinfo{person}{Ed~H. Chi}, {and}
  \bibinfo{person}{Minmin Chen}.} \bibinfo{year}{2024}\natexlab{b}.
\newblock \showarticletitle{LLMs for User Interest Exploration in Large-scale
  Recommendation Systems}. In \bibinfo{booktitle}{\emph{Proceedings of the 18th
  ACM Conference on Recommender Systems}} \emph{(\bibinfo{series}{RecSys
  '24})}. \bibinfo{publisher}{Association for Computing Machinery},
  \bibinfo{address}{New York, NY, USA}, \bibinfo{pages}{872–877}.
\newblock
\showISBNx{9798400705052}
\urldef\tempurl%
\url{https://doi.org/10.1145/3640457.3688161}
\showDOI{\tempurl}


\bibitem[\protect\citeauthoryear{Xu, Hua, and Zhang}{Xu et~al\mbox{.}}{2024}]%
        {xu2024openp5}
\bibfield{author}{\bibinfo{person}{Shuyuan Xu}, \bibinfo{person}{Wenyue Hua},
  {and} \bibinfo{person}{Yongfeng Zhang}.} \bibinfo{year}{2024}\natexlab{}.
\newblock \showarticletitle{Openp5: An open-source platform for developing,
  training, and evaluating llm-based recommender systems}. In
  \bibinfo{booktitle}{\emph{Proceedings of the 47th International ACM SIGIR
  Conference on Research and Development in Information Retrieval}}.
  \bibinfo{pages}{386--394}.
\newblock


\bibitem[\protect\citeauthoryear{Zeng, Yue, Jiang, and Wang}{Zeng
  et~al\mbox{.}}{2024}]%
        {zeng2024federated}
\bibfield{author}{\bibinfo{person}{Huimin Zeng}, \bibinfo{person}{Zhenrui Yue},
  \bibinfo{person}{Qian Jiang}, {and} \bibinfo{person}{Dong Wang}.}
  \bibinfo{year}{2024}\natexlab{}.
\newblock \showarticletitle{Federated recommendation via hybrid retrieval
  augmented generation}. In \bibinfo{booktitle}{\emph{2024 IEEE International
  Conference on Big Data (BigData)}}. IEEE, \bibinfo{pages}{8078--8087}.
\newblock


\end{thebibliography}
\end{document}